\documentclass[pdflatex,sn-mathphys,iicol]{sn-jnl} 

\jyear{2023}%

\theoremstyle{thmstyleone}
\theoremstyle{thmstyletwo}%

\theoremstyle{thmstylethree}%

\usepackage{subcaption}
\usepackage{multirow}
\usepackage{tabularx}
\usepackage{adjustbox}
\usepackage{siunitx}

\newcommand{\pt}{p_\mathrm{T}}
\newcommand{\geant}{G{\footnotesize EANT}4}
\newcommand{\graph}{\mathcal{G}}
\newcommand{\edges}{\mathcal{E}}
\newcommand{\neighborhood}{\mathcal{N}}
\newcommand{\vertices}{\mathcal{V}}
\newcommand{\reals}{\mathbb{R}}
\newcommand{\order}{\mathcal{O}}

\begin{document}

\title[Jet energy calibration with deep learning as a Kubeflow pipeline]{Jet energy calibration with deep learning as a Kubeflow pipeline}

\author[1]{\fnm{Daniel} \sur{Holmberg}}\email{daniel.holmberg@helsinki.fi}
\author[2]{\fnm{Dejan} \sur{Golubovic}}\email{dejan.golubovic@cern.ch}
\author[3]{\fnm{Henning} \sur{Kirschenmann}}\email{henning.kirschenmann@cern.ch}

\affil[1]{\orgdiv{Department of Computer Science}, \orgname{University of Helsinki}, \orgaddress{\postcode{00560 Helsinki}, \country{Finland}}}
\affil[2]{\orgname{CERN}, \orgaddress{\postcode{1211 Geneva 23}, \country{Switzerland}}}
\affil[3]{\orgname{Helsinki Institute of Physics}, \orgaddress{\postcode{00560 Helsinki}, \country{Finland}}}

\abstract{Precise measurements of the energy of jets emerging from particle collisions at the LHC are essential for a vast majority of physics searches at the CMS experiment. In this study, we leverage well-established deep learning models for point clouds and CMS open data to improve the energy calibration of particle jets. To enable production-ready machine learning based jet energy calibration an end-to-end pipeline is built on the Kubeflow cloud platform. The pipeline allowed us to scale up our hyperparameter tuning experiments on cloud resources, and serve optimal models as REST endpoints. We present the results of the parameter tuning process and analyze the performance of the served models in terms of inference time and overhead, providing insights for future work in this direction. The study also demonstrates improvements in both flavor dependence and resolution of the energy response when compared to the standard jet energy corrections baseline.}


\maketitle

\section{Introduction}\label{sec:introduction}

The adoption of machine learning methods has had a profound impact on the field of high energy physics, greatly increasing the discovery potential in the data measured by the particle detectors at the Large Hadron Collider (LHC)~\cite{ml_in_hep}. Deep learning especially has proven very useful~\cite{deep_learning_lhc} with graph neural networks being one of the most expressive and versatile architectures to choose for many tasks~\cite{gnns_in_hep} ranging from reconstructing particle tracks~\cite{track_reconstruction} to classifying complete events~\cite{event_classification}.

In this paper, we study the application of deep learning for calibrating the energy of particle jets at the Compact Muon Solenoid (CMS) experiment~\cite{cms}. Jets in this context originate from high energy proton-proton collisions producing color charged partons that undergo hadronization forming collimated sprays of color neutral particles. Calibrating the energy of jets is an involved process, split into several factorized steps, some based only on simulations and some on comparisons with data~\cite{jec_cms_8tev}. A precise calibration of the jet energy scale is crucial for a wide variety of physics analyses, most prominently e.g. measurements of the top quark mass~\cite{top_mass} and inclusive jet cross-section measurements. 

Deep learning has been successfully applied by the CMS collaboration in the past to increase the energy resolution of bottom jets with a feedforward neural network~\cite{b_jet_regression}. These efforts are here extended to all jet flavors in a QCD-jet data sample publicly accessible on the CERN OpenData portal~\cite{opendata}. Additionally, by adopting recent advancements in representation learning, specifically ones made for jet classification~\cite{pfn, particlenet}, more information about jet constituents can be included in the training process, which has proven beneficial for jet calibration~\cite{efn_calibration}.

Furthermore, since operationalizing machine learning workflows is a challenge in itself for many organizations~\cite{who_needs_mlops}, we introduce a cloud native pipeline for running jet energy calibration experiments. It runs on the Kubeflow platform~\cite{kubeflow_deployment} that comes with readily available components for hyperparameter tuning and model serving among others. Kubeflow has been used by researchers in various domains such as bioinformatics to achieve rapid scaling with containers~\cite{kubeflow_bioinformatics}, or as a means to create automated machine learning workflows for a service-aware 5G network model adapting to drift in the input data~\cite{kubeflow_5G_networks}. Adopting a cloud native workflow enables the workload to be smoothly deployed also on public cloud resources, as was done for fast simulation of electromagnetic showers using generative deep learning at CERN~\cite{kubeflow_gan}.

The rest of this paper is structured as follows. In Section~\ref{sec:data}, we explain the contents of the CMS open dataset used for this study. In Section~\ref{sec:regression}, we go through the data distribution, feature sets and models used to calibrate jet energy. Section~\ref{sec:pipeline} introduces the Kubeflow pipeline used for training and serving our models on internal cloud resources. Section~\ref{sec:results} shows the results that our models yield, and lastly in Section~\ref{sec:conclusion} we draw some final conclusions.

\section{Dataset definition and conventional jet energy calibration}\label{sec:data}

In this study, we utilize a dataset prepared in the context of the CMS OpenData effort~\cite{opendata}. The dataset consists of particle jets extracted from simulated proton-proton ($pp$) collision events at $\sqrt{s} = 13$\,TeV. These events are generated at leading-order perturbative QCD with P{\footnotesize YTHIA}\,8~\cite{pythia}, and include CMS detector simulation and event reconstruction.

``Particle-level jets'' or ``generator jets'' are clustered using the anti-$k_{T}$ algorithm~\cite{anti_kt} with radius parameter of 0.4 from stable (decay length $c\tau > 1$\,cm) final-state particles resulting from the hadronization of partons originating from the $pp$\,collisions. As these particles propagate through the detector, they leave signals in detector components such as the tracker and the electromagnetic and hadronic calorimeter. The modeling of the interaction with the material and detector response relies on the CMS Full simulation, based on \geant~\cite{geant}.

The types of quarks or gluons that initiate the formation of the jet determines the flavor of a jet. Flavor labeling is done using a technique called ``ghost association'' \cite{heavy_flavor}, where heavy flavor hadrons and light quark and gluon partons are added as ``ghost particles'' to the clustering process. If heavy flavor hadron ghosts are found in the jet, it is labeled as a heavy flavor (b or c) jet. If no ghost hadron is found, the jet is checked for light flavor (uds) or gluon (g) partons, identifying it as a corresponding jet \cite{jet_algorithms}.

The Particle Flow (PF) approach at CMS~\cite{particle_flow} is a method that attempts to reconstruct each particle in the event individually, prior to the jet clustering, based on information from all relevant sub-detectors, resulting in a list of ``PF candidates''. Various methods for per-particle pileup  (additional $pp$ collisions occurring in the same bunch crossing as the event of interest) mitigation can be applied~\cite{CMSPUMitigation}. The charged hadron subtraction (CHS) is the default for narrow radius jets in Run\,2 CMS data, meaning that charged particles associated with pileup vertices are removed prior to jet clustering, thereby reducing the impact of pileup on jets. The remaining reconstructed PF candidates are then used as input to  jet clustering algorithms to form a ``reconstructed jet'' (anti-$k_{T}$, R=0.4) that is supposed to be as close to the particle-level jet in terms of kinematic quantities as possible. 

The aim of jet energy corrections is to correct the energy of the reconstructed jets---on average---back to the jet energy of particle-level jets. The measured jet energy is affected by  various effects, such as energy loss in the detector material, non-linear response of the detector, and pileup. CMS factorizes the jet energy corrections into levels to individually correct for various effects~\cite{jec_cms_8tev}: The L1 correction corrects for offset energy induced by pileup. It is determined from simulated samples with/without pileup, parametrizing the offset energy as a function of median event energy density, jet area, $\pt$, and $\eta$. The L2L3 correction corrects for remaining detector response dependence and is parametrized as a function of jet $\pt$ and $\eta$. The L2L3Residual correction corrects for any remaining differences between experimental data and simulation, but is not applicable in this MC-only study. 

These conventional jet energy corrections do not take into account the substructure of the jets. The PF approach already leads to a reduced dependence of, e.g. the jet response on the flavor of a jet in comparison to calorimeter-only reconstruction. However, taking the substructure of jets into account promises potential for reduced flavor response differences and associated uncertainties as well as an improved jet-energy resolution.

\section{Jet energy regression}\label{sec:regression}

Standard jet energy corrections can be further improved upon by using supervised machine learning. An important step in this direction has been taken by the CMS Collaboration for b jets specifically~\cite{b_jet_regression}. For these jets a significant part of the jet energy is carried by semileptonically decaying b-hadrons, decaying to charged leptons and neutrinos. Neutrinos escape detection by the CMS detector since they only interact via the weak force leading to jet energy being underestimated. However, using a neural network trained on a sample of simulated top quarks event decaying into b jets and W bosons significant improvements in the energy resolution for b jets were achieved.

These efforts can be generalized to other jet flavors too. Using a QCD sample such as the one described in Section~\ref{sec:data} enables the training of a regression model for multiple jet flavors. This approach can potentially address any flavor discrepancies in the energy response. Notably, the response of light quark jets and gluon jets can vary significantly due to the higher color charge of gluon jets, which results in more and softer particles with a lower calorimeter response. Although the PF algorithm can reduce the response difference substantially by replacing the non-linear calorimeter measurement of charged hadron energy with the corresponding track momentum, 15\% of jet energy is still carried by neutral hadrons subject to the calorimeter response non-linearity~\cite{particle_flow}.

\subsection{Data distribution and input features}

\begin{figure}[h]
    \centering
    \includegraphics[width=\columnwidth]{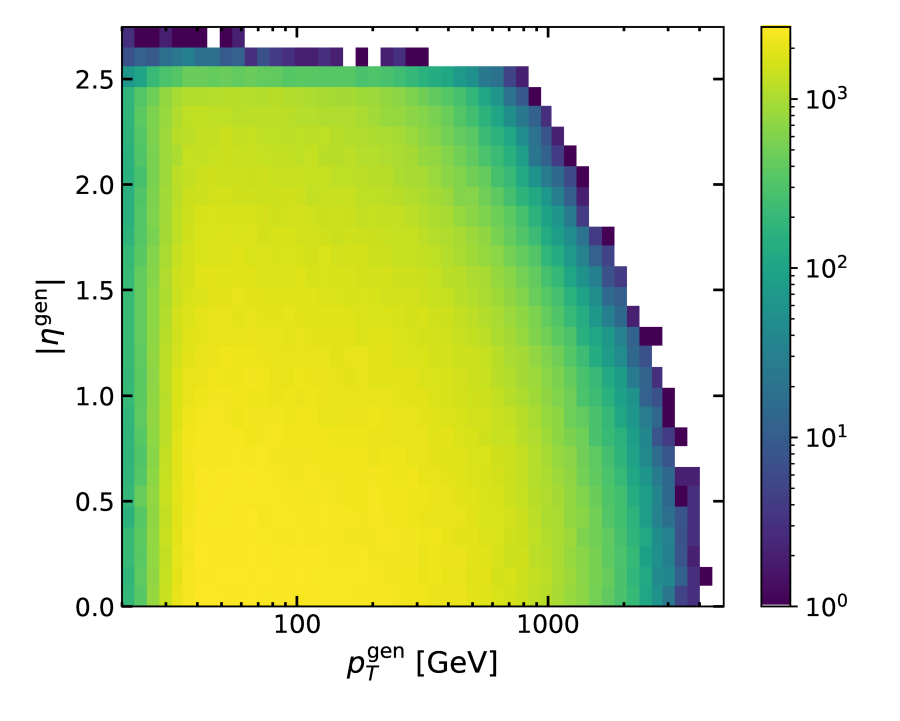}
    \caption{A heatmap illustrating the distribution of jets in the dataset with respect to the generated transverse momentum ($\pt^\text{gen}$) and the absolute value of generated pseudorapidity ($\lvert \eta^\text{gen} \rvert$)}
    \label{fig:data}
\end{figure}

The jets in the CMS open dataset span a large spectrum both in terms of $\pt$ and pseudorapidity $\eta$ as seen in Figure~\ref{fig:data}. However, the very low $\pt$ region experiences high pileup, resulting in lower-quality jets. Additionally, the forward region of the detector lacks tracking information leading to a worse event reconstruction quality and less reliable measurements to train on. To address these issues, the dataset is filtered by $\pt^\text{gen} \gt 20$\,GeV and $\lvert \eta \rvert \lt 2.5$. In total 1.42M jets were used, 60\% of which were allocated to the training set. The model was validated at the end of every training epoch on a separate set with 20\% of all jets. Once the training had finished, model performance was evaluated on a test set with the remaining 20\% of all jets.

Table~\ref{tab:features} presents the features used for training our regression model selected from a comprehensive list of variables available in the CMS open dataset~\cite{opendata}. Eight of the features describe jets as a whole. The reconstructed $\pt$ is log-transformed to reduce the width of the otherwise long-tailed distribution. In addition to $\pt$, jet coordinates $\eta$ and $\phi$ as well as jet mass and catchment area are also included in the set of training features. The last three jet-level variables $\pt D$, $\sigma_2$ and multiplicity can help to discriminate between quark and gluon jets~\cite{qgl}.

Every particle within a jet has six associated features. The two most statistically significant for the regression task at hand are: 1) the log-transformed $p_{T_i}$ of a PF candidate indexed with $i$, and 2) the same variable, but relative to the $\pt$ of the whole jet, also log-transformed. Additionally, each PF candidate has four location-based features: the $\eta_i$, $\phi_i$ and $\theta_i$ detector coordinates, along with the distance $R_i$ from the particle to the center of the jet. 

\begin{table*}[ht]
    \centering
    \caption{Overview of the input features used for training the regression model, categorized into jet-level features and PF candidate features. Jet features describe the overall properties of the jet, while PF features characterize the individual particles within the jet}
    \label{tab:features}
    \begin{tabular}{cll}
    \bottomrule
    \midrule
    \textbf{Category} & \textbf{Variable} & \textbf{Description}                     \\ 
    \midrule
    \multirow{8}{*}{Jet features}
    & $\log \pt$                    & Logarithm of a jet's $\pt$           \\ 
    & $\eta$                        & Pseudorapidity of a jet                \\ 
    & $\phi$                        & Azimuthal angle of a jet               \\ 
    & $m$                           & Mass of a jet                                 \\ 
    & $A$                           & Catchment area of a jet                       \\ 
    & $\pt D$                       & Fragmentation distribution for a jet's $\pt$ \\ 
    & $\sigma_2$                    & Minor ellipse axis of a jet \\ 
    & multiplicity                  & Jet constituent multiplicity             \\ 
    \midrule
    \multirow{6}{*}{PF features}
    & $\log p_{T_i}$                  & Logarithm of a particle's $\pt$ \\
    & $\log \frac{p_{T_i}}{\pt}$      & Logarithm of the fractional $\pt$ of a particle     \\
    & $\eta_i$                        & Pseudorapidity of a particle               \\ 
    & $\phi_i$                        & Azimuthal angle of a particle               \\ 
    & $\theta_i$                      & Polar angle of a particle                   \\ 
    & $R_i$                           & Distance from a particle to the center of the jet     \\ 
    \bottomrule
    \bottomrule
    \end{tabular}
\end{table*}

\subsection{Regression target and loss function}

\begin{figure}[h]
    \centering
    \includegraphics[width=\columnwidth]{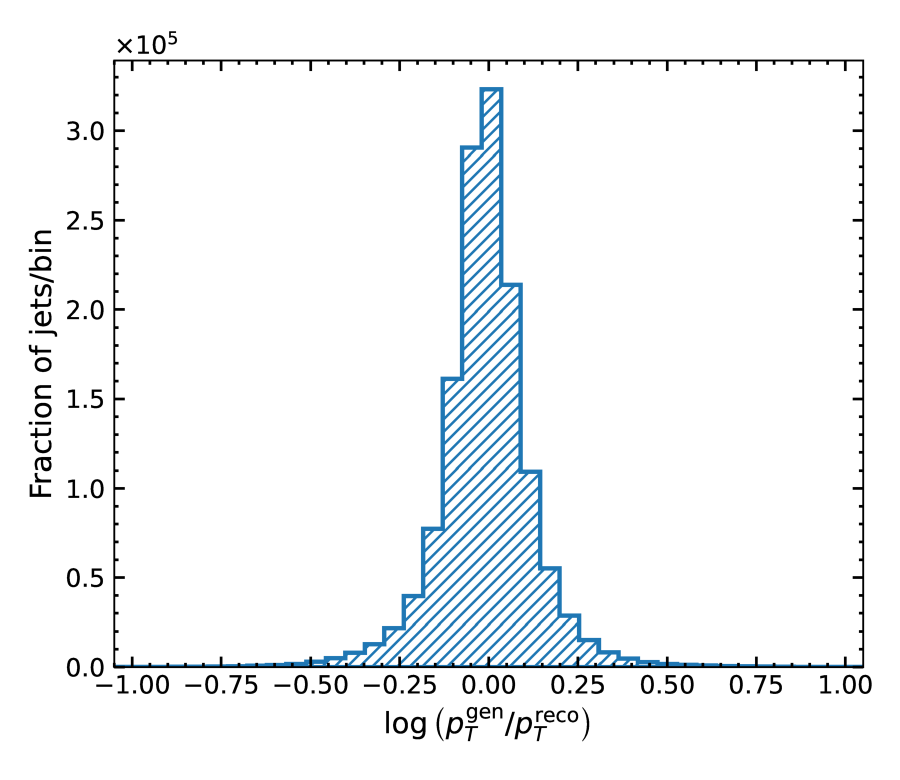}
    \caption{Distribution of the regression target}
    \label{fig:target}
\end{figure}

The aim of the regression is to train a model to map a a set of features describing a reconstructed jet towards the transverse momentum of the corresponding generator-level jet. However, by itself the particle-level $\pt^\text{gen}$ follows a decreasing exponential distribution that covers many orders of magnitude in the energy spectrum as seen in Figure~\ref{fig:data}. To counteract this, $\pt^\text{gen}$ can be divided by the similarly distributed $\pt^\text{reco}$ that is part of the training set. This gives a target distribution on the order of one with a reduced variance compared to the original target distribution~\cite{b_jet_regression}. To further correlate the target with the input features the logarithm is taken yielding the final regression target as $y = \log\left( \pt^\text{gen} / \pt^\text{reco} \right)$. The distribution of the target is narrow and centralized around zero as seen in Figure~\ref{fig:target}. In order to get the corrected transverse momentum $\pt^\text{corr}$, the exponential function is applied to the prediction $\hat{y}$, and the resulting correction factor is multiplied by $\pt^\text{reco}$. The per-jet energy response can then be defined as $R = \pt^\text{corr} / \pt^\text{gen}$.

The mean absolute error (MAE) is selected as the loss function to minimize for this problem, as it assigns less importance to outliers compared to the more commonly used mean squared error (MSE) loss. In line with the previous study on b jet energy regression~\cite{b_jet_regression}, a loss function with reduced sensitivity to the tails of the target distribution is preferred. To prevent potential spikes in the training loss, jets with a target value smaller than -1 or larger than 1 are excluded, as they are considered too poorly reconstructed to be taken into account.  The loss function used is thus:
\begin{equation}
    \label{eq:loss}
    L = \frac{1}{N} \sum_{i=1}^N \lvert y_i - \hat{y}_i \rvert I_{\lvert y_i \rvert \lt 1}.
\end{equation}

A further motivation for the use of the MAE loss function is that the statistic it learns is the median of the target distribution, whereas for example the minimum of the MSE loss lies on the function that maps input features to the expected value of the target. Predicting the median in this case can be seen as beneficial since it is a robust measure of central tendency. By learning the median of the target distribution, the model is more likely to make accurate predictions even in the presence of outliers.

\subsection{Deep learning models}\label{sec:models}

During recent years many new approaches of applying deep learning in jet physics have emerged, especially for the purpose of jet tagging where the aim is to classify jets based on the particle initiating them. Some proposed approaches to do this is for example to treat the jets as images~\cite{jet_images}, sequences~\cite{jet_sequences} or trees~\cite{jet_trees}. While these methods perform well and surpass traditional multivariate methods, the way they represent jet constituents is not ideal. A particle jet can contain up to $\order(100)$ particles whereas an image of a jet will contain $\order(1000)$ pixels, and as noted in~\cite{jet_images} the images are indeed very sparse with 5-10\% of pixels being active. When instead considering a sequence or tree of particles as the representation a notable issue that arises is that the particles must be ordered in some fashion to be used by the deep learning model, e.g. a recurrent neural network or recursive neural network. However, the constituents of a particle jet have no intrinsic ordering.

More natural ways of representing particles have been found by adopting point cloud based formalisms from the wider machine learning community as particle clouds~\cite{particlenet}. This kind of representation treats a collection of particles as a graph structure $\graph = (\vertices, \edges)$ where each individual particle serves as a node $\vertices = \{1, \dots, n\}$ with potential edges $\edges \subseteq \vertices \times \vertices$ connecting them. In this work, we compare two separate models that operate on data represented in this way.

The simplest case of representation learning on particle clouds is when the set of edges is empty $\edges = \varnothing$. This will lead to the particles taking on the form of an unordered set, and was originally proposed for the Particle Flow Network (PFN) model~\cite{pfn} adapting from the Deep Sets~\cite{deep_sets} framework. The key idea here is that input feature vectors $\mathbf{x}_i \in \reals^F$ are mapped with an equivariant function into a latent feature vector $\mathbf{h}_i=\psi(\mathbf{x}_i)$. In practice that would be a multilayer perceptron (MLP) with shared weights for all elements of the set. 

\begin{figure*}[ht]
    \centering
    \includegraphics[width=0.748\textwidth]{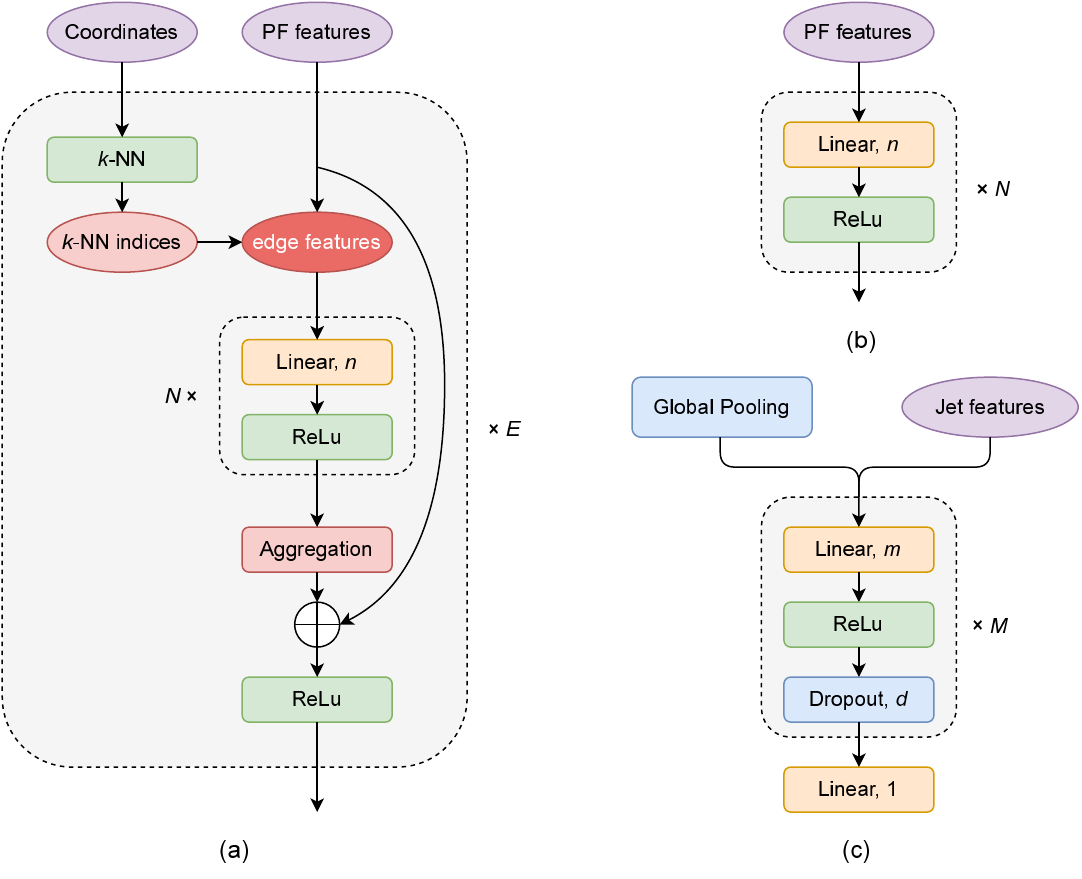}
    \caption{Illustration of model architectures: (a) EdgeConv block for ParticleNet, (b) Deep Sets block for PFN, and (c) Network head shared by both models}
    \label{fig:model_architectures}
\end{figure*}

To make predictions for the particle jet as a whole the latent feature vectors must be aggregated using a permutation invariant pooling, such as summing, averaging or taking the max value. Following the PFN and Deep Sets papers, summation $\sum_{i \in \vertices}$ is chosen as the global pooling operation. Figure~\ref{fig:model_architectures}\,(b) in conjunction with Figure~\ref{fig:model_architectures}\,(c) show the complete network architecture where the output of the Deep Sets block connects to global pooling in the network head. The global particle representation is concatenated with jet features before being passed into a final MLP mapping towards the regression target. Note that the rectified linear unit (ReLU)~\cite{relu} is used as activation function, and dropout~\cite{dropout} is applied in the head of the network.

Spatial information can be used to further increase the expressivity of a point cloud based model. The ParticleNet~\cite{particlenet} architecture uses edge convolution (EdgeConv), first introduced as a building block of dynamic graph convolutional neural networks (DGCNN)~\cite{dgcnn}, to incorporate information on the local neighborhood of each particle. Detector coordinates in the ($\eta$, $\phi$)-plane are used to calculate the Euclidean distance matrix from pairwise distances between particles. The $k$-NN algorithm is then applied to construct edges connecting each particle to its $k$ nearest neighboring particles.

Messages between every point $\mathbf{x}_i$ and its neighbors $\mathbf{x}_j$ are learned using an asymmetric edge function $\psi( \mathbf{x}_i , \mathbf{x}_j - \mathbf{x}_i )$ implemented as an MLP with shared weights. Permutation invariant aggregation in the form of averaging $\frac{1}{k} \sum_{ i \in \neighborhood_i^k }$ over the learned edge features for the $k$ nearest neighbors is used to update every node in the particle cloud. A shortcut connection~\cite{shortcut} from the original node features is added to the output of the aggregation before passed being passed through the ReLU activation function. This concludes the EdgeConv block shown in Figure~\ref{fig:model_architectures}\,(a).

If multiple EdgeConv blocks are stacked after one another the input graphs are dynamically updated by calculating the pairwise distance matrix from the latent feature space learned by the previous block. Global average pooling $\frac{1}{n} \sum_{i \in \vertices}$ is applied on the output of the last EdgeConv block as it is passed into the network head in Figure~\ref{fig:model_architectures}\,(c). An identical procedure to that in the PFN model is applied, where the jet features are concatenated with the pooled particle features, and finally passed through one last MLP mapping towards the regression target.

The models are implemented in PyTorch~\cite{pytorch} as part of the weaver deep learning framework for high energy physics~\cite{weaver}. Weaver supports handling common particle physics data formats such as ROOT~\cite{root} or Awkward Array~\cite{awkward}, as well as distributed training, and model inference. To scale up the training on cluster resources where GPUs are distributed over separate nodes the code must support collective communications to synchronize gradients over machines. Different backends can be chosen in PyTorch for this purpose such as Message Passing Interface (MPI)~\cite{mpi} for CPU parallelization, or NVIDIA Collective Communication Library (NCCL)~\cite{nccl} for multi-GPU communication.

\section{Kubeflow pipeline}\label{sec:pipeline}

The analysis was carried out on the Kubeflow-based machine learning platform at CERN~\cite{kubeflow_deployment}. Kubeflow is an open-source machine learning toolkit that supports the entire machine learning lifecycle by providing features such as notebooks, pipelines, hyperparameter optimization, distributed training, model serving, and model monitoring. Kubeflow is built on top of Kubernetes~\cite{kubernetes}, a container orchestrator, leveraging the scalability, ease of use and integration of cutting-edge infrastructure technologies. Deployed as a set of Kubernetes resources, Kubeflow application code runs as a collection of micro-services that communicate with each other and process user workloads.

\begin{figure*}[ht]
    \centering
    \includegraphics[width=\textwidth]{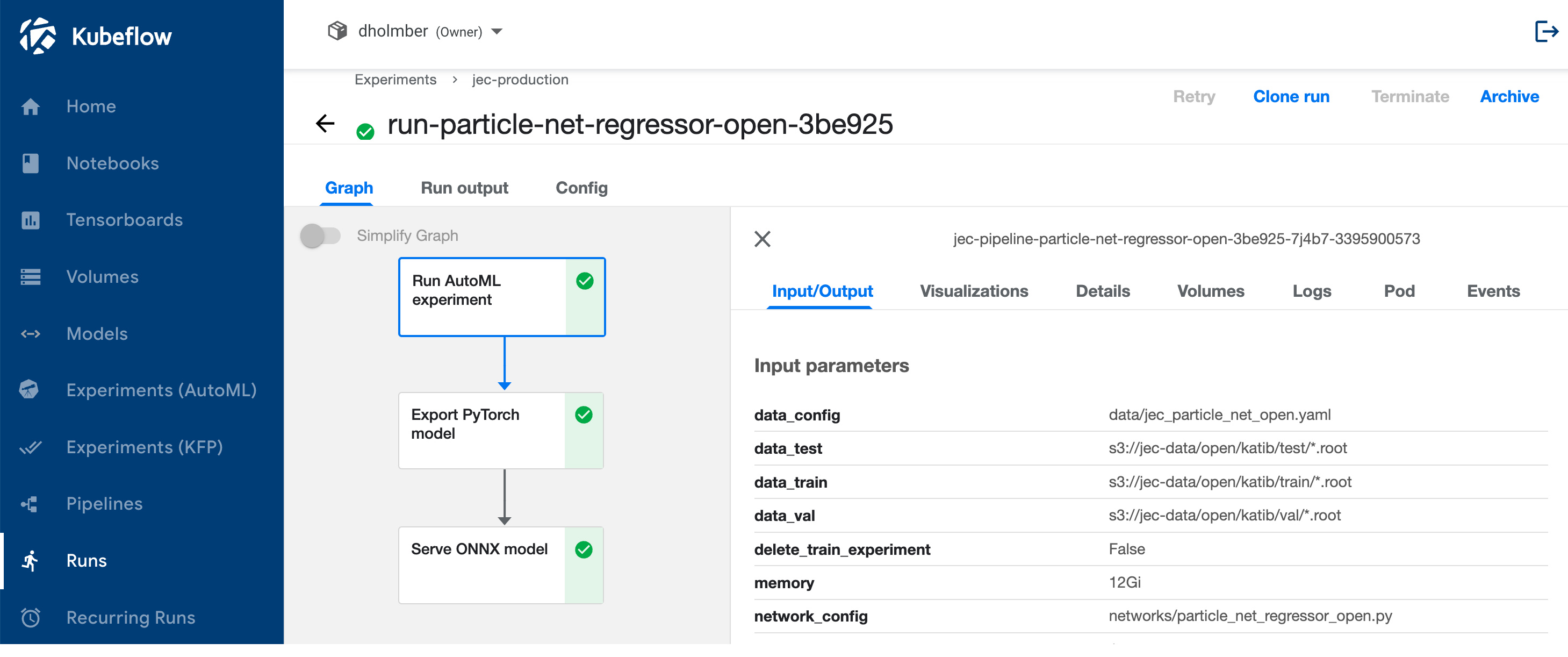}
    \caption{Kubeflow UI for a jet energy regression pipeline run. The pipeline consists of three steps: 1. hyperparameter tuning using Kubeflow's AutoML component Katib, 2. exporting the optimal PyTorch model to the ONNX format, and 3. serve the exported model over HTTP with KServe}
    \label{fig:kubeflow}
\end{figure*}

Machine learning workflows typically take the form of a directed acyclic graph that begins with data processing, proceeds through model training, and ends with an inference phase of the trained model. Kubeflow facilitates developing such workflows by offering several features, including a web interface for managing, tracking, and running pipelines, an engine for scheduling pipeline steps, a software development kit (SDK) for defining and running pipelines using Python, and higher-level abstraction tools like KALE~\cite{kale}, which convert notebooks to pipelines. Regardless of how a pipeline is defined, it is always converted to a Kubernetes YAML~\cite{yaml} definition file before being submitted for execution. A pipeline runs as a sequence of pods (containers), with each pipeline step waiting for its dependencies to complete successfully before proceeding. The pipeline developed for this study is shown as part of the Kubeflow user interface (UI) in Figure~\ref{fig:kubeflow}.

Pipelines offer benefits in resource utilization. By allowing users to define hardware requirements for each step, pipelines ensure that GPUs are only utilized when needed, for example during training and inference steps. Other steps that use CPU-only resources make GPUs available for other users in the cluster. Additional features include support for pipeline scheduling which enables automatic execution of repeated or periodic workflows, running pipelines with different input parameters without any code changes, and grouping pipeline runs into experiments making it easier to track and compare similar runs. 

\subsection{AutoML experiment}\label{sec:hp_tuning}

The Kubeflow Katib component~\cite{katib} offers a streamlined process for automated machine learning (AutoML) supporting hyperparameter tuning, early stopping and neural architecture search. Here we use hyper-parameter optimization, that includes three main steps: 1. implementing a script that trains a model and takes hyperparameters as command line arguments, 2. building a docker image with all dependencies to run the script, and 3. specifying a YAML file with the definition of the hyper-parameters. The YAML file defines the search algorithm, early stopping options, maximum number of parallel jobs, hardware resources, and other options. The YAML file can be generated manually, using an SDK, or using the high-level KALE tool.

\begin{figure}[h]
    \centering
    \includegraphics[width=0.89\columnwidth]{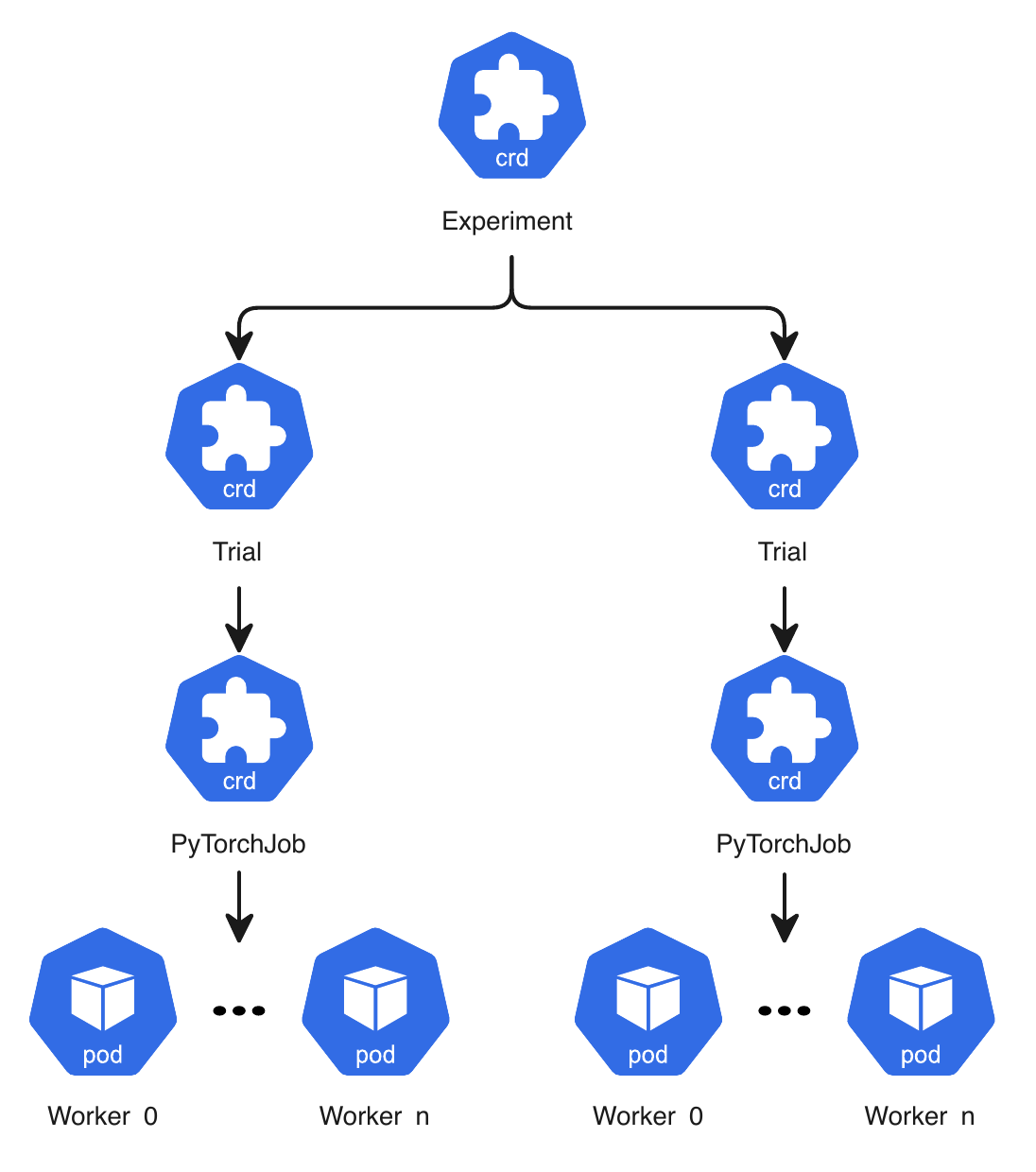}
    \caption{The structure of an Experiment Kubernetes custom resource. An AutoML experiment consists of multiple trials, with each trial representing a unique combination of hyperparameters. Each trial monitors a PyTorchJob, which can submit one or more workers to train the model. Once completed, the Experiment resource retains the outcomes of all trials and the optimal trial result based on predefined metrics}
    \label{fig:katib}
\end{figure}

Katib schedules search jobs in the form of multiple trials, each trial corresponding to a unique combination of hyperparameters as seen in Figure~\ref{fig:katib}. A trial can be a Kubernetes job running a script in a single pod, a pipeline where multiple pods run sequentially or a distributed training job where multiple pods run synchronously utilizing multiple GPUs to train the model. Being framework-agnostic, Kubeflow supports running search jobs with any machine learning framework.

In this study, Katib trials were executed using the PyTorchJob Kubernetes custom resource~\cite{pytorchjob}. The Kubeflow training operator component provides distribution on the level of containerization; training a single model using multiple cluster GPUs that can be located in different machines. In addition to implementing the training to use distribution strategies, it is necessary to specify a YAML definition of a distributed job. The YAML definition includes many attributes, including the number of worker replicas, memory, and command line arguments to the machine learning code. The PyTorchJob also includes S3 credentials to access the training data stored in a bucket on CERN object storage.

To optimize the performance of ParticleNet and PFN, Katib was configured to use the Random Search algorithm~\cite{random_search} to find the optimal set of parameters for minimizing the test loss. The total number of trials was set to 60 for both models with 10 trials running in parallel using one GPU each to achieve fast scheduling of pods on the cluster. Each trial ran for 50 epochs, with the batch size set to 500. The initial learning rate is set as part of the hyperparameter optimization process. After 70\% of epochs, 35 in this case, a scheduler starts decreasing the learning rate exponentially on a per-epoch basis down to 1\% of the initial value at the end of training. The model with the lowest validation loss is chosen as the final model for each trial, and is then run on the evaluation set to get the test loss that we are trying to minimize.

With reference to Figure~\ref{fig:model_architectures}\,(b) and \ref{fig:model_architectures}\,(c), the search space for the hyperparameter tuning was defined as follows for PFN:
\begin{itemize}
    \item linear layers: $N \in \{1, 2, 3, 4, 5\}$
    \item linear layer units: $n \in \{50, 100, 200, 400\}$
    \item linear layers: $M \in \{1, 2, 3, 4, 5\}$
    \item linear layer units: $m \in \{50, 100, 200, 400\}$.
\end{itemize}
The search space for ParticleNet, with respect to Figure~\ref{fig:model_architectures}\,(a) and \ref{fig:model_architectures}\,(c), was defined as:
\begin{itemize}
    \item EdgeConv blocks: $E \in \{1, 2, 3\}$
    \item nearest neighbors: $k \in \{4, 8, 16\}$
    \item linear layers: $N \in \{1, 2, 3\}$
    \item linear layer units: $n \in \{50, 100, 200\}$
    \item linear layers: $M \in \{1, 2, 3\}$
    \item linear layer units: $m \in \{50, 100, 200\}$.
\end{itemize}
Lastly, the mutual hyperparameters considered were:
\begin{itemize}
    \item dropout rate: $d \in [0; 0.5]$
    \item initial learning rate: $lr \in [10^{-5}; 10^{-2}]$
    \item optimizer: $optim \in$ \{AdaGrad, Adam, AdamW, Ranger, RMSProp\}.
\end{itemize}

The completion of trials can be viewed directly from the Katib UI, while Kubeflow's TensorBoard component allows tracking the progress of individual training runs. To set up a TensorBoard server, it is necessary to specify a model output path, either on a persistent volume claim (PVC) within the cluster or an S3 object storage endpoint.  As long as the model training is writing to the specified location, the model performance can be monitored in real-time using TensorBoard servers.

\subsection{Exporting the optimal model}

Once the optimal hyperparameters for both PFN and ParticleNet models have been determined through the hyperparameter tuning process, the optimal PyTorch model is exported to the ML framework agnostic Open Neural Network Exchange (ONNX) format~\cite{onnxruntime}. This enables seamless integration with other ML tools, such as NVIDIA Triton Inference Server~\cite{triton} for model serving, and eases the deployment process.

The second step of the Kubeflow pipeline involves running a PyTorchJob to carry out this conversion. It retrieves the optimal PFN and ParticleNet models from the S3 bucket, converts them to ONNX format, and stores the resulting ONNX models back into the same S3 bucket. The PyTorchJob can be defined using a YAML file that specifies the necessary configurations, such as the PyTorch model input path and the ONNX model output path, network configuration, and hardware resources to run the export job.

When exporting a model to ONNX, a configuration file is created alongside the model file to facilitate serving the model using Triton. We made a schema in the Protocol Buffers (protobuf)~\cite{protobuf} format with model input/output dimensions, data types (32-bit float) and graph optimization level for ONNX Runtime. This was compiled into a Python file that can be used to automatically generate model configuration in protobuf text format with the correct input and output dimensions when exporting a model in PyTorch.

ONNX Runtime defines a static computational graph for the model as opposed to the dynamic one used by PyTorch during training, which allows for various graph optimizations that can improve inference performance, such as graph-level transformations, node eliminations, node fusions, and layout optimizations. An extended optimization level is available that enables complex node fusions. However, these optimizations were found to cause issues when serving ParticleNet, and as a result they were only applied to PFN. A more basic graph optimization level with semantics-preserving graph rewrites that removes redundant nodes and redundant computation was chosen for ParticleNet.

To allow Triton to accept dynamically varying batch sizes, we configured the maximum batch size to be 100k in the model configuration file. Batch requests that large are not necessarily recommended due to the spiky network load they would produce and the excessive amount of memory that must be allocated to the inference server.

\begin{figure}[h]
    \centering
    \includegraphics[width=0.635\columnwidth]{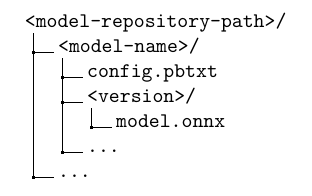}
    \caption{Model repository layout for Triton Inference Server with ONNX backend}
    \label{fig:model_repository}
\end{figure}

The export job produces an output directory structure, as shown in Figure~\ref{fig:model_repository}, that follows Triton's specifications. The base model repository is in our case an S3 bucket path and a unique id for every pipeline run. The top-level repository can contain many subdirectories (or pseudo-folders since object storage has a flat address space), each representing distinct models. The optimal ONNX model is placed in a numeric sub-folder signifying model version during exportation, and the automatically generated model configuration file is placed alongside that folder.

\subsection{Model serving}

After exporting the optimal PFN and ParticleNet PyTorch models to the ONNX format and storing them in an S3 bucket, the models are served using custom InferenceService resources. An InferenceService is the interface used for deploying models on Kubeflow's inference platform KServe~\cite{kserve}. 

\begin{figure*}[ht]
    \centering
    \includegraphics[width=0.8\textwidth]{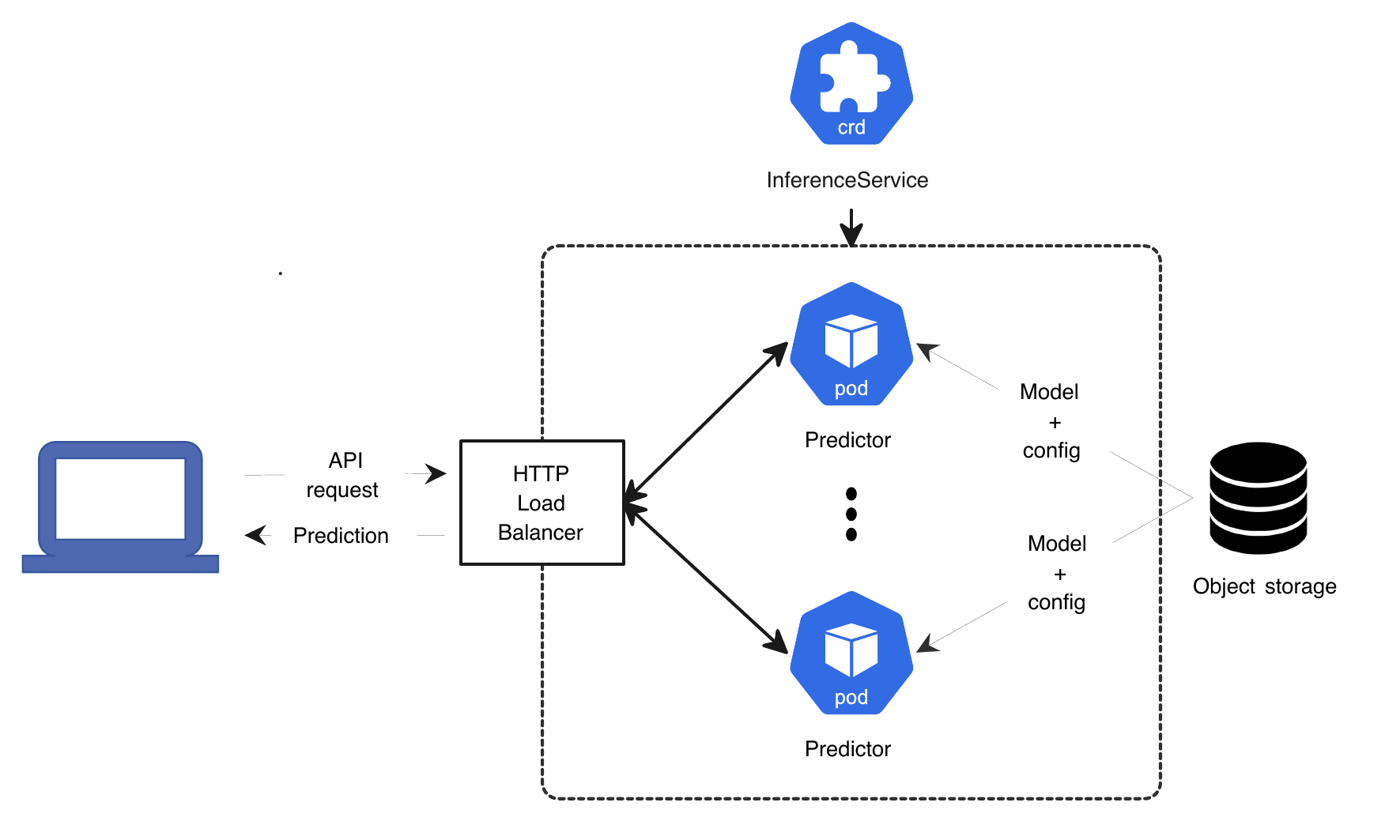}
    \caption{A diagram depicting model serving using KServe, highlighting load balancing of user inference requests and the scalability of predictor pods}
    \label{fig:serving}
\end{figure*}

The InferenceService can be specified using a YAML file with various configuration options, such as hardware allocation for the server (CPU, GPU and memory), runtime version of the predictor, and the path to the model repository. In order to make authenticated requests to S3 storage, a Kubernetes ServiceAccount with the required access rights was deployed on the cluster and attached to every InferenceService.

KServe relies on Knative~\cite{knative} for scaling serverless workloads and supports scale-to-zero, optimizing cost efficiency. Istio~\cite{istio}, another key technology in KServe, acts as a service mesh that uses Kubernetes sidecars (containers deployed alongside a main container in a pod) for network traffic management, providing features such as progressive ``canary'' model rollouts, traffic routing, ingress management, logging, load balancing, and security.

Triton is utilized as the predictor for the InferenceService. It is an open-source inference server capable of serving multiple models concurrently, supporting various machine learning frameworks. We have specified the ONNX Runtime as platform in the model configuration file telling Triton explicitly which backend to use.

The complete inference workflow is illustrated in Figure~\ref{fig:serving}. The InferenceService creates Triton pods that retrieves the ONNX model and configuration from S3 object storage. The pods act as REST endpoints \cite{rest} that can be queried over HTTP. When user sends inference requests, a load balancer is responsible for receiving them and distribute them to available inference server pods. The pods can scale up or down dynamically based on the volume of incoming requests. Servers pass the input data through the deep learning model and return model output to the load balancer that sends it back to the user.

\section{Results}\label{sec:results}

\subsection{Hyperparameter optimization}

The results of the hyperparameter tuning using Random Search can be analyzed to some extent with Pearson's correlation coefficient. Table~\ref{tab:correlation} shows the correlation of all continuous and ordinal hyperparameters with the inverted test loss for the models with the lowest validation loss during each training run. Because the choice of hyperparameters is stochastic for every trial it is difficult to isolate the impact of a single hyperparameter on the model's performance, and thus particularly high correlation scores are not expected here. Note also that the correlation is limited to the search space laid out in Section~\ref{sec:hp_tuning}.

\begin{table*}[h]
\centering
\caption{A two-fold presentation of the results from hyperparameter tuning. The upper part of the table shows the Pearson correlation between hyperparameters and the inverted test loss for PFN and ParticleNet. The bottom part lists the top three best sets of hyperparameters found for both models with the corresponding test loss values for those trials. Gain signifies the relative improvement in loss for each trial compared to the standard corrections baseline ($L_\text{Baseline}=9.427$e-2), and is computed as $\Delta L / L_\text{Model}$}
\label{tab:correlation}
\begin{adjustbox}{width=\textwidth}
\begin{tabular}{llccccccccccc}
    \bottomrule
    \midrule
    & Model & $E$ & $k$ & $N$ & $n$ & $M$ & $m$ & $d$ & $lr$ & $optim$ & $L$ & Gain [\%] \\
    \midrule
    \multirow{2}{*}{$\rho_{\mathrm{hp}, 1/L}$}
    & PFN         &            &            & 0.27       & -0.21      & -0.08      & 0.33       & -0.54      & -0.39 & & &  \\
    & ParticleNet & 0.22       & -0.04      & 0.17       & -0.21      & 0.02       & 0.40       & -0.39      & -0.34 & & &  \\
    \midrule
    \multirow{6}{*}{top trials}
    & \multirow{3}{*}{PFN}
    & & & 5 & 50 & 3 & 400 & 9.45e-3 & 1.08e-3 & Ranger & 8.768e-2 & 7.52 \\
    & & & & 5 & 50 & 4 & 100 & 4.13e-2 & 7.60e-4 & Adam & 8.770e-2 & 7.49 \\
    & & & & 3 & 200 & 4 & 400 & 8.58e-2 & 7.33e-4 & RMSProp & 8.770e-2 & 7.49 \\
    & \multirow{3}{*}{ParticleNet}
    & 3 & 16 & 3 & 50 & 3 & 200 & 1.16e-2 & 8.72e-3 & Ranger & 8.746e-2 & 7.79 \\
    & & 3 & 16 & 2 & 100 & 2 & 100 & 1.14e-1 & 1.84e-3 & Ranger & 8.752e-2 & 7.71 \\
    & & 3 & 8 & 2 & 50 & 3 & 200 & 8.70e-2 & 2.50e-3 & AdamW & 8.755e-2 & 7.68 \\
    \bottomrule
    \bottomrule
\end{tabular}
\end{adjustbox}
\end{table*}

The table suggests that for PFN, having more linear layers with fewer units in the Deep Sets block is weakly associated with a lower loss. For ParticleNet, having more EdgeConv blocks with additional linear layers and fewer units offers an advantage. The number of nearest neighbors $k$ in ParticleNet's particle graph appears uncorrelated with a lower loss. The remaining correlation values exhibit similar behavior for both models. The loss is relatively unaffected by the number of linear layers in the network head, but more units in these layers tend to yield a lower loss. Increased dropout negatively impacts the regression task with the most certainty among all correlation results. Finally, a lower learning rate tends to produce better results.

The best hyperparameters found for both models, as listed in the lower section of Table~\ref{tab:correlation}, do tend to align with the Pearson correlation scores. For the PFN model, the top three trials all used a configuration with more linear layers (3--5) in the Deep Sets block and fewer units (50--200), which aligns with the correlation scores observed for these parameters. Similarly for ParticleNet, the top three trials incorporate more EdgeConv blocks (3) with a higher number of linear layers (2--3) and fewer units (50--100), reflecting the corresponding correlations. The initial learning rate and dropout across the top trials are, with exception for the optimal ParticleNet trial's learning rate, notably low as suggested by the Pearson correlation.

To further highlight the impact of poorly adjusted dropout and learning rate we can compare the average of those parameters for trials that fall in the upper and lower quartiles ranked by test loss. For PFN, the average initial learning rate for trials in the lower quartile of losses was 3.0e-3, while for trials in the upper quartile it was higher, at 5.6e-3. Similarly, the average dropout for trials in the lower quartile was 0.11, compared to a significantly higher 0.31 in the upper quartile. A similar pattern was observed with the ParticleNet model, with the average initial learning rate for trials in the lower quartile being 3.3e-3, compared to 6.3e-3 in the upper quartile, and the average dropout for trials in the lower quartile being 0.16, compared to 0.31 in the upper quartile. 

A high learning rate allows the model to learn quickly, but it may also cause the model to overshoot the optimal solution and not converge well. The aim of dropout on the other hand is for the model to learn more robust, generalizable representations of the data. However, if the dropout rate is too high, as is the case for many trials in the upper loss quartile, the model may struggle to learn from the data at all, leading to underfitting.

The optimizer algorithm as a nominal variable falls outside the scope of the correlation analysis. However, Ranger proved to be the most successful. It combines LookAhead~\cite{lookahead} with $k = 6$ and $\alpha = 0.5$, and an inner RAdam optimizer~\cite{radam} with $\beta_1 = 0.95$, $\beta_2 = 0.999$ and $\epsilon = 10^{-5}$.  RAdam can help to stabilize the learning rate and adapt it based on the variance of the gradient, making it a robust option when the learning rate is ill-adjusted. Furthermore, LookAhead has empirically been shown to improve convergence by considering multiple directions in the parameter space. It also mitigates the impact of poorly chosen hyperparameters on training by smoothing out noisy gradient updates.

The gain measure in Table~\ref{tab:correlation} represents the percentage improvement in test loss achieved by the PFN and ParticleNet models over the standard jet energy corrections loss value. This metric serves as an indicator of the models' relative performance, providing a quantifiable measure of the benefits realized through hyperparameter optimization. The gain observed for the optimal PFN configuration is 7.52\% whereas the optimal ParticleNet model achieves a 7.79\% improvement over the baseline.

In this study, Random Search proved straightforward to set up and it showcased favorable practical properties. Random state is the only input parameter, the algorithm allows for trials to be discontinued or restarted without jeopardizing the experiment, and compared to grid search, it is more efficient for a given computational budget~\cite{random_search}. However, Katib offers several other AutoML algorithms such as Bayesian optimization~\cite{bayesian} or Hyperband~\cite{hyperband} that could be considered for future work since they have great potential to more effectively find an optimal set of hyperparameters.

It should be noted that more advanced algorithms often rely on additional input parameters that may alter the outcome which adds a level of complexity in the setup. Furthermore, while Random Search is embarrasingly parallel, Bayesian Optimization uses Gaussian process regression to iteratively model the search space and is therefore inherently sequential. Trials can still be queued in parallel, but the choice of parameter configuration for back-to-back trials is less informed than when the algorithm is run sequentially. Hyperband, on the other hand, is an extension of Random Search, and offers more efficient resource allocation to trials that matter by invoking early stopping on poorly performing configurations. However, adjusting resource allocation when candidate configurations have different convergence rates is an open challenge~\cite{hyperband}, a circumstance occurring in our experiment with varying learning rates and models with differing numbers of layers and hidden units.

\subsection{Model complexity and inference performance}

Table~\ref{tab:model_complexity} compares the complexity of the optimal PFN and ParticleNet models in terms of loss, number of parameters, and Multiply-Accumulate operations (MACs). MACs represent the number of multiplications and additions performed during a single forward pass, indicating computational complexity. While ParticleNet achieves a lower test loss than PFN due to the inclusion of particle locality information, it has significantly higher computational complexity. A smaller number of nearest neighbors in the particle graph and fewer channels in the linear layers can be considered for reducing the complexity while still maintaining good performance~\cite{particlenet}.

\begin{table}[h]
\caption{Comparison of model complexity for the optimal configuration of PFN and ParticleNet}
\label{tab:model_complexity}
\centering
\begin{tabular}{lccc}
    \bottomrule
    \midrule
    Model & Loss & \# Params & MACs \\
    \midrule
    PFN & 8.768e-2 & 355.45k & 1.43M \\
    ParticleNet & 8.746e-2 & 123.47k & 47.59M \\
    \bottomrule
    \bottomrule
\end{tabular}
\end{table}

\begin{figure*}[ht]
    \centering
    \includegraphics[width=\textwidth]{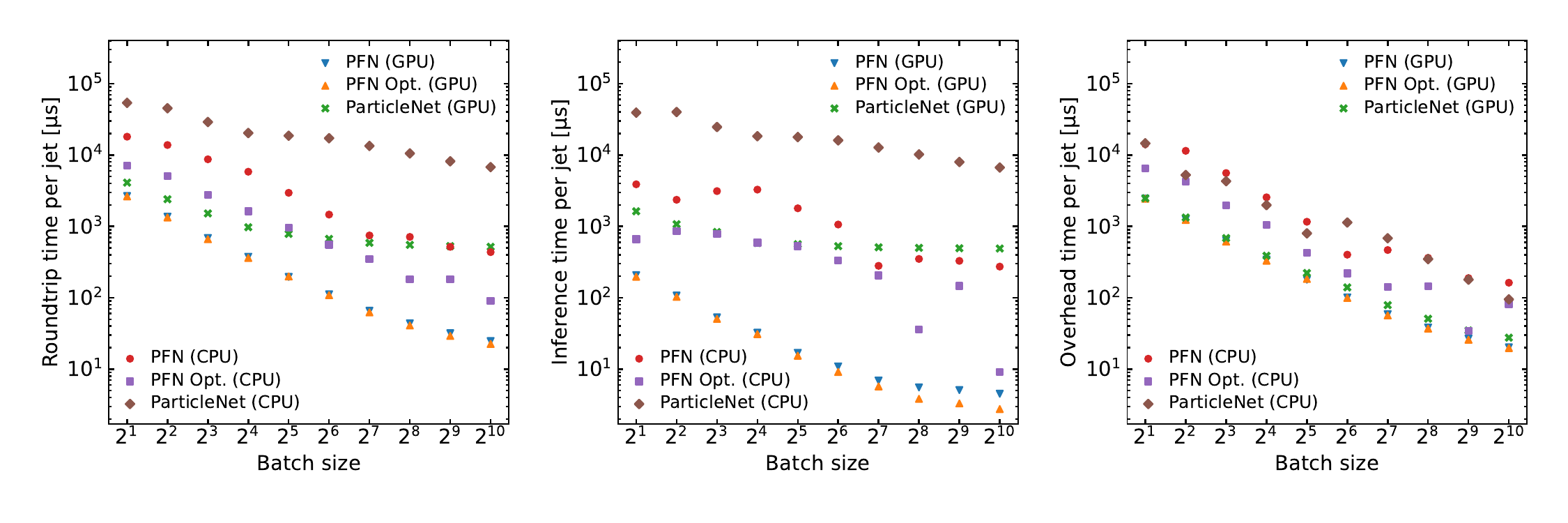}
    \caption{Comparison of prediction request roundtrip time (left), device inference time (middle) and overhead (right) for PFN and ParticleNet ONNX models served with Triton. The represented values are based on the average processing time for 1k repetitions with randomly selected jets, covering a range of batch sizes from 2 to 1024}
    \label{fig:throughput}
\end{figure*}

The optimal PFN and ParticleNet models are served as REST endpoints using the Triton Inference Server running on top of KServe. We used the Python Triton client to request predictions for different batch sizes to evaluate how these models compare, and how request batch size affects roundtrip time, inference time, and overhead. The roundtrip time encompasses the total duration for a request to be processed, including both HTTP request time and inference time. Meanwhile, the overhead, calculated as the difference between roundtrip time and inference time, represents additional delay induced by factors such as data serialization / deserialization and network latency. The results of these tests for both models when served either on a CPU or a Tesla V100 GPU are displayed in Figure~\ref{fig:throughput}.

In the context of model performance, PFN achieve lower roundtrip times than ParticleNet due to its much lower computational complexity. However, the difference is less pronounced for smaller batch sizes. The larger overhead at small batch sizes affects both models similarly, thereby reducing the relative performance difference between the models. 

For very fast models such as PFN running on GPU the inference time is minimal due to the low model complexity and the high parallel processing capabilities of the GPU. This leads to overhead being the dominating factor in the roundtrip time. Conversely, for slower models like ParticleNet, the inference time especially when processing large batch sizes is substantially longer than the overhead time. As a result, the overhead comprises a fraction of the roundtrip time in that scenario.

We can note that PFN with extended ONNX graph optimization (PFN Opt.) shows performance enhancements compared to PFN with basic graph optimizations. Extended graph optimizations go beyond the simple, semantics-preserving transformations of basic optimizations, and applies complex node fusions after graph partitioning, tailoring the computation to the specific execution provider (CPU or GPU). This results in more efficient computations that are better suited to the architecture of the hardware on which the model runs.

The faster inference time on GPU compared to CPU for both PFN and ParticleNet models can be primarily attributed to the difference in the underlying architecture of these hardware platforms. GPUs are specifically designed for high-throughput, parallel processing and are capable of executing thousands of threads simultaneously. This characteristic is particularly beneficial for inference tasks which involve large-scale matrix operations that can be parallelized effectively. In contrast, CPUs have fewer cores and are optimized for sequential tasks. Furthermore, the performance gap can be widened when the batch size is large, as larger batch sizes enable better utilization of the GPU's parallel processing capabilities, leading to a faster per-jet inference time.

GPU inference and overhead times are also more consistent compared to CPU times. When allocating a processor in a Kubernetes cluster the instance is assigned a virtual processor (vCPU) shared across different processes, and due to their general-purpose nature, CPUs are typically tasked with managing a wider variety of processes, including system operations and other applications running in the background. A scheduler managed by the kernel has to coordinate time slots on the physical CPUs, which can lead to more variability in the availability of resources for the inference tasks. In contrast, GPU allocations on Kubeflow will currently result in a dedicated GPU for the task, resulting in more consistent performance. 

When faced with the choice between CPU and GPU for deep learning inference it really depends on the hardware resources available and application requirements. Hardware accelerators such as GPUs provide superior throughput but often come with increased costs and power consumption. For applications with moderate inference workloads and less stringent response time requirements, CPU-based inference may be more cost-effective. In contrast, high-throughput, low-latency applications can benefit significantly from GPU investment. It is also noteworthy that while a dedicated GPU will yield good results when benchmarking, virtualization of these resources as vGPUs could lead to better resource utilization~\cite{vgpu_cern}, which is especially important as demand for hardware accelerators continues to increase.

\subsection{Jet energy response flavor dependence}

\begin{figure*}[h]
    \centering
    \includegraphics[width=\textwidth]{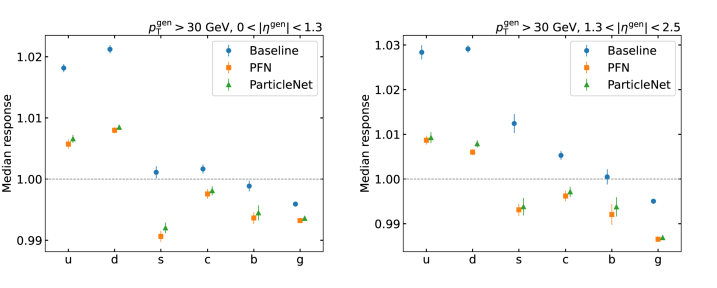}
    \caption{Median energy response separated by jet flavor in the barrel region (left) and the endcap region (right). The improvement over the baseline is about 40\% in the barrel region and 50\% in the endcap region of the detector for both models}
    \label{fig:flavor_comparison}
\end{figure*}

In this section, we present the analysis of the flavor dependence of the calibrated energy response produced by the optimal PFN and ParticleNet models. Figure~\ref{fig:flavor_comparison} displays the median response for each jet flavor obtained using the two deep learning models and the standard corrections baseline. Both models exhibit a reduction in the differences between jet flavors compared to the baseline. A notable improvement in flavor dependence of the energy calibration is observed between light quark jets and gluon jets, which has been a known shortcoming in standard jet energy corrections.

The uncertainty for each data point, represented by the error bars in Figure~\ref{fig:flavor_comparison}, is calculated using the statistical bootstrapping method. By randomly sampling all response values 30 times, new sets of response values are generated similar in magnitude to the original dataset. The median is computed for each sample, followed by the standard deviation of the 30 median values, which provides an uncertainty for each point. There are fewer s, c and b jets in the QCD sample compared to the amount of u, d and g jets contributing to them having a higher bootstrapped uncertainty. The sample also has fewer jets in the endcap region compared to barrel region resulting in a higher uncertainty for the median response in the right plot in Figure~\ref{fig:flavor_comparison} compared to the left one.

The sum of absolute errors (SAE) is used to evaluate the improvement in flavor dependence. It can be computed directly from the points in Figure~\ref{fig:flavor_comparison} by summing the absolute difference between the median response for each flavor and the mean of the same points. Mathematically, it can be expressed as:
\begin{equation}
    \mathrm{SAE} = \sum_\mathrm{flavor} \lvert R_{50\%,\: \mathrm{flavor}} - \frac{1}{n}\sum_\mathrm{flavor} (R_{50\%,\:\mathrm{flavor}}) \rvert
\end{equation}
where $\mathrm{flavor}=\{u, d, s, c, b, g\}$ represents the complete set of jet flavors in the QCD sample. The relative improvement in flavor dependence for a model compared to the standard JEC is denoted by $\alpha$ and is defined as:
\begin{equation}
    \alpha = 1 - \mathrm{SAE}_{\mathrm{Model}} / \mathrm{SAE}_{\mathrm{Baseline}} .
\end{equation}
The values that $\alpha$ can take on ranges from any negative value to one, where a negative value indicates that the model performs worse than the standard correction, zero corresponds to no improvement, and $\alpha = 1$ would mean that the median response produced by the deep learning model is identical for all jet flavors.

As seen in Table~\ref{tab:results}, the improvement in flavor dependence varies across different $\pt$ intervals for both models. In the very low $\pt$ region ($30\,\mathrm{GeV} < \pt^\mathrm{gen} < 100\,\mathrm{GeV}$), the improvements in flavor dependence are modest. As the $\pt$ intervals become larger, both models demonstrate more significant improvements. Which of the two models perform better in a certain $\pt$ interval or detector region varies. However, the differences in performance between them become less pronounced as $\pt$ increases. It is also worth noting that the improvement in flavor dependence is not uniform across the barrel and endcap regions. For both models, the improvement is generally larger in the endcap region, especially in the intermediate and high $\pt$ intervals.

\begin{table*}[h]
    \centering
    \caption{Summary of jet energy regression results. Improvements in flavor dependence and energy resolution produced by PFN and ParticleNet compared to baseline JEC are presented in multiple $\pt$ intervals and detector regions}
    \label{tab:results}
    \begin{tabular}{@{}clcccc@{}}
    \bottomrule
    \midrule
        Interval [GeV] & Model & $\alpha_\mathrm{barrel}$ [\%]    & $\beta_\mathrm{barrel}$ [\%] & $\alpha_\mathrm{endcap}$ [\%] & $\beta_\mathrm{endcap}$ [\%] \\
    \midrule
    \multirow{2}{*}{$30 < \pt^\mathrm{gen} < 100$}
        & PFN           & 8.04 & 0.23 & 12.01 & 0.97 \\
        & ParticleNet   & 6.71 & 0.64 & 17.97 & 1.25 \\
    \midrule
    \multirow{2}{*}{$100 < \pt^\mathrm{gen} < 300$}
        & PFN           & 23.25 & 1.43 & 49.59 & 6.93 \\
        & ParticleNet   & 24.52 & 1.61 & 45.07 & 7.07  \\
    \midrule
    \multirow{2}{*}{$300 < \pt^\mathrm{gen} < 1000$}
        & PFN           & 57.52 & 4.53 & 68.05 & 12.90 \\
        & ParticleNet   & 56.11 & 4.81 & 70.02 & 11.93 \\
    \midrule
    \multirow{2}{*}{$\pt^\mathrm{gen} > 1000$}
        & PFN           & 68.62 & 7.95 & 37.91 & 4.97 \\
        & ParticleNet   & 68.34 & 7.75 & 37.56 & 9.37 \\
    \bottomrule
    \bottomrule
    \end{tabular}
\end{table*}

Flavor response differences and even more so their differences between different generators are an input to both ATLAS and CMS flavor-related uncertainties. If the flavor responses become more alike, because the underlying jet properties are taken into account, as demonstrated in Table \ref{tab:results} and Figure \ref{fig:flavor_comparison}, then one can also expect the uncertainties based on generator differences to be decreased.

\subsection{Jet energy resolution}

\begin{figure*}[h]
    \centering
    \includegraphics[width=\textwidth]{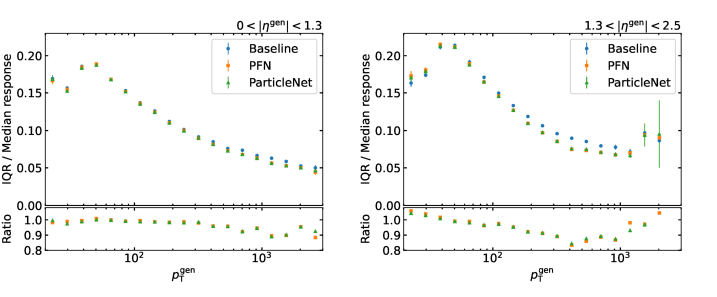}
    \caption{Relative jet energy resolution in the barrel region (left) and the endcap region (right) binned logarithmically. The bottom panels show the ratio between the relative resolution produced by the deep learning models and the relative resolution after standard jet energy corrections}
    \label{fig:resolution}
\end{figure*}

The performance of the regression models can also be assessed by examining the relative jet energy resolution. We define it here as the interquartile range (IQR) divided by the median for the response:
\begin{equation}
    \bar{s} = \frac{R_{75\%} - R_{25\%}}{R_{50\%}} .
\end{equation}
The IQR serves as a measure of response resolution. Both the median and IQR are robust statistics, meaning that they are less affected by outliers compared to the mean and standard deviation respectively. The uncertainty of the relative resolution is measured using the same bootstrapping technique as in the previous section and results are shown in Figure~\ref{fig:resolution}. As the high $\pt$ and endcap region are less populated as indicated in Figure~\ref{fig:data}, the uncertainties are sizeable for endcap high $\pt$ jets.

The improvement in relative resolution for a model with respect to standard corrections is denoted here as $\beta$. We define it as one minus the ratio of relative jet energy resolution between the models and the baseline:
\begin{equation}
    \beta = 1 - \bar{s}_{\mathrm{Model}} / \bar{s}_{\mathrm{Baseline}} .
\end{equation}

When examining the results presented in Table~\ref{tab:results}, we observe that both PFN and ParticleNet models achieve improvements in the energy resolution compared to the baseline. The improvements vary across different $\pt$ intervals, with the largest improvements observed in the intermediate to high $\pt$ intervals for both models. This behavior can be attributed to several factors. High $\pt$ jets generally have more complex substructures and are more likely to undergo hard parton splittings, resulting in a higher multiplicity for the jet. This increased complexity leaves more room for improvement for machine learning-based approaches. The effect of pileup also diminishes at higher $\pt$ resulting in less noisy data to train on. Owing to the limited amount of training data in the endcap region for jets with $\pt^\text{gen} > 1000$\,GeV and reaching the kinematic limit of phase space in that regime, the improvements achieved through deep learning are comparatively smaller there.

\section{Conclusion}\label{sec:conclusion}

In this paper, we presented a deep learning based workflow for calibrating the energy of particle jets in the CMS detector. By utilizing advancements in learning on particle clouds in the form of the PFN and ParticleNet models, we managed to improve upon standard jet energy corrections derived solely from kinematic quantities. The results, categorized into jet energy resolution and flavor dependence, suggest that the performance of both networks is generally comparable, with larger improvements at higher $\pt$. The most notable difference between the two models is that the inclusion of locality information in ParticleNet results in a slightly better energy resolution at the expense of higher model complexity and inference time.

We have also demonstrated the potential of the Kubeflow platform for operationalizing ML workflows in high energy physics. As the field is witnessing a growing integration of ML techniques, the capabilities offered by Kubeflow, supporting the continual development of scalable ML solutions, are becoming increasingly more relevant. The pipeline we developed for this work enabled us to efficiently scale up our AutoML experiments on cloud resources and serve the optimal models as easily queryable REST endpoints. Having each step in the pipeline defined using Kubernetes custom resources allows for fine-grained access to hardware resources on the cloud and well-versioned, reusable machine learning workflows. 

\bmhead{Acknowledgments}

We wish to thank the CMS collaboration and the CERN OpenData group for publishing high-quality simulated data under an open-access policy. We acknowledge the support of the CERN IT department for providing the computational resources for this work. 

\bmhead{Funding}

Corresponding author D.H. is supported by the Academy of Finland under the ICT 2023: Frontier AI Technologies program (Grant No. 345635).

\bmhead{Data availability statement}

The simulated dataset used for this study is hosted on the CERN OpenData portal. The instructions and code to replicate the studies in this paper are available at: \url{https://zenodo.org/record/7799179}

\bibliography{sn-article} 


\end{document}